\documentclass[%
 reprint,
 amsmath,amssymb,
 aps,
prb,
]{revtex4-1}	

\usepackage{graphicx}
\usepackage{dcolumn}
\usepackage{bm}

\begin{document}


\title{Interplay between $p-$ and $d-$ orbitals yields multiple Dirac states in one- and two-dimensional CrB$_4$ }

\author{Alejandro Lopez-Bezanilla$^{1}$}
\email[]{alejandrolb@gmail.com}
\affiliation{Theoretical Division, Los Alamos National Laboratory, Los Alamos, NM 87545, United States}

\date{\today}
             
\begin{abstract}

Theoretical evidence of the existence of six inequivalent and six threefold degenerate pairs of Dirac cones in the low-spectrum diagram of monolayered hexagonal CrB$_4$ is provided.  
The four $d$ electrons of the Cr atom are yielded to the B sublattices creating an isoelectronic structure to graphene where the interplay between $p$ and $d$ orbitals leads to the appearance of additional Dirac states on both one- and two-dimensional geometries.
{\it Ab initio} calculations show that, although spin-orbit interaction splits the cone-shaped valence and conduction bands, CrB$_4$ is a semimetal with compensated electron-hole pockets.
As the two-dimensional layer is shaped into finite-width ribbons, one actual and one symmetry-frustrated Dirac point are observed at the Fermi level, yielding massless fermions in a one-dimensional nano-structure with no topological insulating features.  
A rational explanation in terms of periodic boundary conditions across the ribbon axis is given to unveil the origin of the Dirac point.

\end{abstract}

\maketitle
\section{Introduction}
With the goal of integrating tomorrow's electronic devices, two-dimensional (2D) materials hosting pseudo-relativistic massless Dirac fermions have been proposed as promising alternatives for post-silicon electronics. Achieving some of the most compelling features from a device perspective, namely high electron-mobility and low-power dissipation, Dirac materials are characterized by a low-energy spectrum where the energy of the electrons is a linear function of their speed. 
Experimental realizations have confirmed the ability of group IV 2D crystals (graphene\cite{Novoselov666}, silicene\cite{PhysRevLett.108.155501,2053-1583-3-1-012001}, germanene\cite{germanene14,0953-8984-27-44-443002}, stanene\cite{stanene15}) for hosting Dirac states. Simultaneously, a surge of interest in predicting new compounds assisted by computational approaches led to the description of materials (graphynes\cite{PhysRevLett.108.086804}, 8-Pmmn borophene\cite{PhysRevLett.112.085502,PhysRevB.93.241405}) with relativistic 1/2 spin particles governed by the Dirac equation.

Dirac fermions emerge at isolated points in the Brillouin zone (BZ), where the top of the valence band and the bottom of the conduction band meet. In buckled 2D arrangements of group-IV atoms, spin-orbit coupling (SOC) leads to a meV-large band gap opening. In all cases, Dirac cones are associated to atoms arranged in a hexagonal network that may extend in a 2D geometry.
Prominent among the many challenges in finding a 2D material featuring Dirac states is to branch away from the single-atom compound, and combine the electronic properties of individual atoms so that linear dispersion at the vicinity of the Fermi energy can emerge. Binary compositions usually yield gapped electronic structures often as a consequence of the ionic character of the chemical bonds (BN\cite{doi:10.1021/nn301675f}) or of the quantum confinement (MoS$_2$\cite{PhysRevLett.105.136805}). The progress made on the synthesis and chemistry of all-B monolayers opens new avenues since they may provide a fundamental template to build new nanostructures with tailored properties\cite{zrb2,PhysRevB.90.161402}.

\begin{figure}[htp]
 \centering
      \includegraphics[width=0.5 \textwidth]{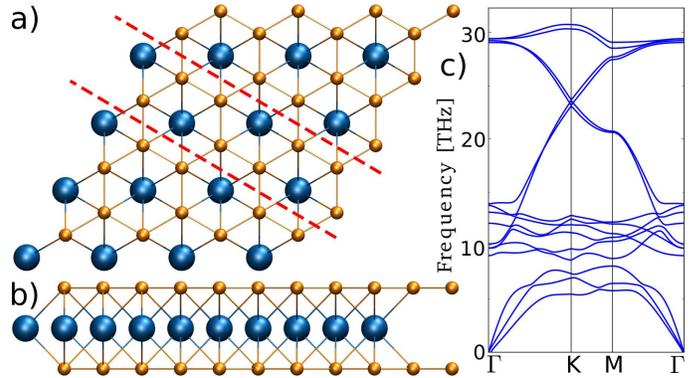}
 \caption{Top a) and b) side view of a schematic representation of the hexagonal cell of monolayered CrB$_4$. Cr atoms sit in between two hexagon centers of two parallel B honeycomb lattices. Dashed lines point out the cut along which a ribbon as shown in Fig.\ref{fig5} is constructed. c) Phonon spectrum of monolayered CrB$_4$}  
 \label{fig1}
\end{figure}

In this paper, theoretical evidence of the existence of Dirac cones in 2D monolayered CrB$_4$ and one-dimensional (1D) nanoribbons is presented. Monolayered CrB$_4$ can be regarded as a hybrid nanostructure composed of a triangular network of Cr atoms in between two B honeycombed lattices (see Fig.\ref{fig1} (a) and (b)). 
Although the three valence electrons of a B atom are efficient in creating sp$^2$ hybrid orbitals that overlap with neighbouring B atoms to create a hexagonal network, a B sublattice considered independently is energetically not viable. However, as explained in ref.[\onlinecite{PhysRevLett.99.115501}]: "{\it this sheet is highly prone to accepting electrons to increase its stability should they be available from another source}". 

\begin{figure}[htp]
  \centering
  \includegraphics[width=0.5  \textwidth]{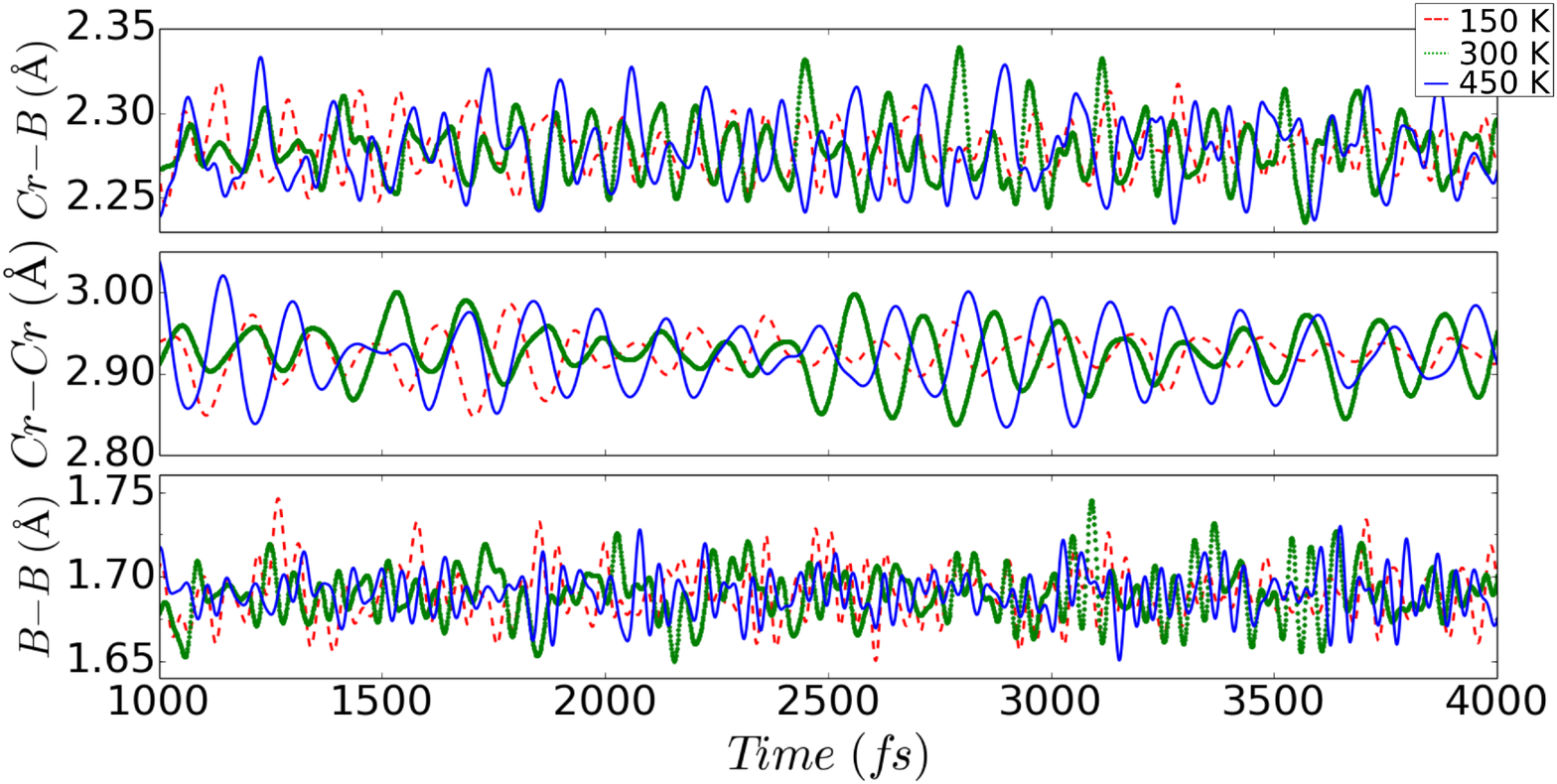}
 \caption{  Molecular dynamic simulations at 150 K, 300K, and 450 K show evidence of the structural stability of two-dimensional CrB$_4$ beyond the harmonic approximation at 0 K. Average bond length variations between pairs Cr-B atoms (upper panel), Cr-Cr atoms (middle panel), and B-B atoms in the same honeycombed lattice (lower panel) remain constant throughout 3 ps after an equilibration time of 1 ps. }
 \label{figMD}
\end{figure}

\begin{figure*}[htp]
 \centering
  \includegraphics[width=0.99 \textwidth]{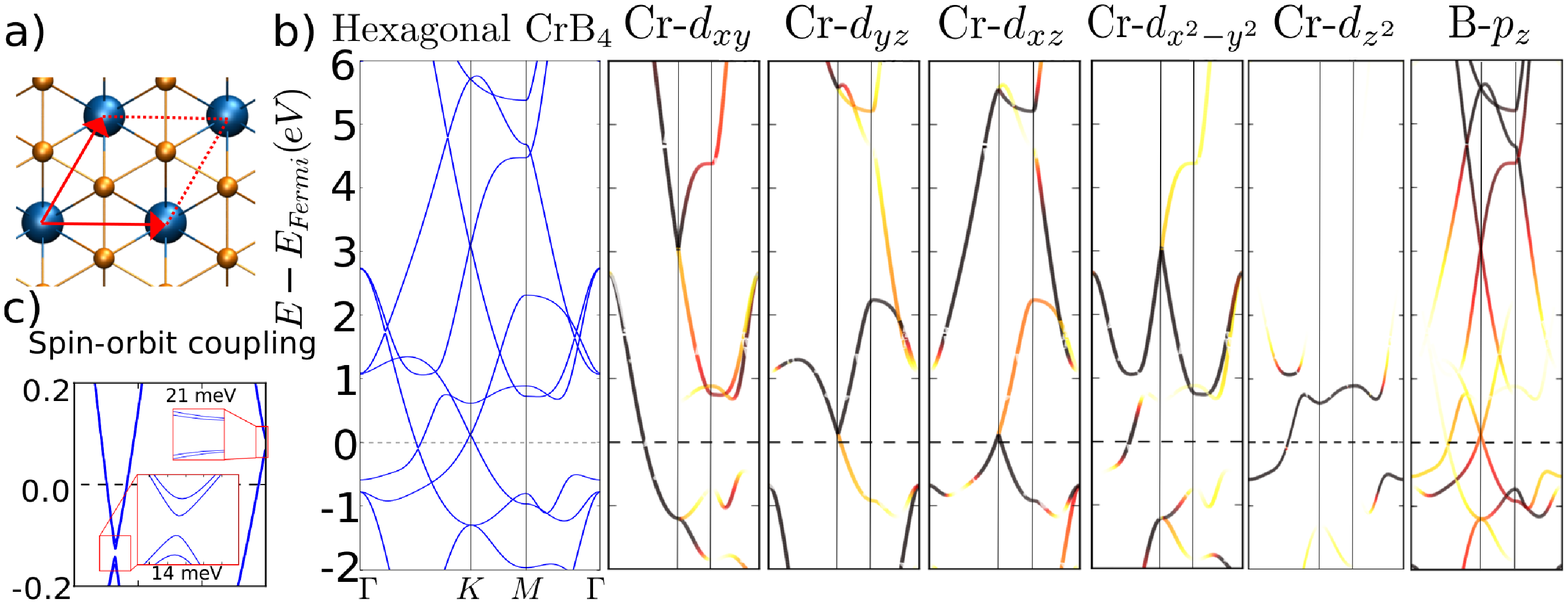}
 \caption{a) Hexagonal unit cell of CrB$_4$. b) Electronic band diagram of CrB$_4$. Two Dirac cones at the Fermi level exhibit compensated electron and hole pockets. Color-resolved density of states show the participation of each atomic orbital to the electronic states of CrB$_4$. Vertical lines point out the position of the high symmetry points as indicated in the $\Gamma \rightarrow K \rightarrow M \rightarrow \Gamma$ path. c) Calculations including spin-orbit coupling demonstrate that Dirac cones are separated by a tiny gap conserving the electron-hole pockets. Horizontal dashed lines point out the Fermi level. 
}
 \label{fig2}
\end{figure*}

The theoretical chemical bonding analysis and self-doping approach proposed by Tang and Ismail-Beigi\cite{PhysRevB.80.134113} achieved experimental realization in the work by Mannix et al.\cite{Mannix1513}, where an additional B atom located on the center of the hexagon provides the exact amount of electrons to stabilize the resulting triangular lattice. External 3$d^2$ and 4$d^2$ transition metal atoms sitting on top of the hexagonal rings demonstrated to yield stable two dimensional monolayered ZrB$_2$\cite{zrb2} and TiB$_2$ \cite{PhysRevB.90.161402}. Following this strategy, if two honeycomb B layers are simultaneously in contact with an additional triangular lattice of atoms located in between, the latter must own twice as much of electrons than in the case of Zr and Ti diborides. The four electrons transfered from the 3$d^4$ Cr orbitals to both B sublattices allows CrB$_4$ to become a stable hybrid structure. The interplay between Cr $d$- and B $p$-orbitals allows for the creation of delocalized orbitals featuring Dirac electronic states. 

In this study, {\it ab initio} calculations show that monolayered CrB$_4$ hosts an unusual number of Dirac states in the vicinity of the Fermi level, both in 1D and 2D geometries. In the BZ of the extended sheet six inequivalent and six threefold degenerate pairs of Dirac cones are formed with compensated electron-hole pockets. Shaping the 2D sheet into a 1D zigzagged ribbon, an actual Dirac point generated by the p$_z$-orbitals of B sublattice coexists with a frustrated Dirac point resulting from the overlap of in-plane Cr $d$-orbitals. In all studied ribbons and regardless of their width, both quantum confinement and SOC are inefficient in vanishing the Dirac point. Boundary conditions on the structured ribbons, charge transfer from Cr to B atoms, and electronic decoupling between sublattices are responsible for the stability of monolayered CrB$_4$ and the formation of Dirac states.

\section{Computational Methods}
Calculations based on the density functional theory (DFT) were performed within the Perdew-Burke-Ernzerhof (PBE) generalized gradient approximation (GGA) functional for the exchange correlation, and the projector-augmented-wave method as implemented in VASP\cite{PhysRevB.48.13115,PhysRevB.54.11169,PhysRevB.59.1758}. The electronic wave functions were computed with plane waves up to a kinetic-energy cutoff of 500 eV. The integration in the 2D (1D) k-space was performed using a 128$\times$128$\times$1 (128$\times$1$\times$1) Monkhorst-Pack k-point mesh centered at the $\Gamma$-point. Atomic coordinates and lattice constants were fully relaxed until the residual forces were smaller than 1 meV/\AA. For an accurate determination of the Fermi energy location, single-energy point calculations were performed with up to 300 k-points in the directions where periodic boundary conditions apply. The force-constant method and the PHONOPY package \cite{phonopy} were employed for computing phonon spectra.
Molecular dynamic simulation were performed with the SIESTA \cite{PhysRevB.53.R10441,0953-8984-14-11-302} code to guarantee the structural integrity and robustness of the nano-structure beyond the harmonic approximation.

 \section{Results and Discussion} 
Coatings of chromium borides are known to possess high hardness, chemical inertness, high thermal and electrical  conductivity. Experimental synthesis of crystals of chromium CrB$_2$ have been reported, where extremely dense, defect-free, crystalline structures of CrB$_2$ were deposited by pulsed magnetron sputtering of loosely packed blended powder targets\cite{AUDRONIS20051366,1742-6596-26-1-086}. A controlled sputtering hot-pressed process or appropriate sintered ceramic targets selection may allow for a precise thickness of the boride coating deposition.
Here we focus on the {\it P6/mmm} point group monolayered phase of CrB$_2$, where the unit cell contains a hexagonal lattice of B atoms with Cr atoms on top of the hexagons. A second honeycombed B layer is placed parallel to the first one over the triangular Cr lattice. Thus, the fully relaxed structure of the freestanding single-layer sheet of CrB$_4$ (Fig.\ref{fig1}a and b) is composed of one atom of Cr sitting in between the center of two hexagons of two parallel honeycombed boron lattices. The lattice parameter of the hexagonal lattice (Fig.\ref{fig2}a) is 2.89 \AA\ and the Cr-B distance is 2.24 \AA. The thickness of the structure is 3.0 \AA.

Dynamical stability is demonstrated by the absence of imaginary branches in the phonon spectrum shown in Fig.\ref{fig1}c. The relatively low dispersion range of $\sim$ 9 THz of the acoustic branches contrasts with the spanning over 20 THz of the optical branches. 
Experimental evidence demonstrates that boron-based nano-structures may be additionally stabilized by interacting with an appropriate substrate. Thus, all-B monolayers were synthesized on Ag \cite{Mannix1513,FengB} enabling the formation of borophenes with several geometries and compositions stabilized by dispersion interactions.

Although it is known that bulk CrB$_2$ is a stable compound at room temperature \cite{PhysRevB.90.064414}, it is worth verifying the robustness of monolayer CrB$_4$ at finite temperature. DFT-based molecular dynamic simulations with a Nos\'e thermostat were performed. Molecular dynamics simulations were performed in the NVT ensemble for a 5$\times$5$\times$1 unit supercell using only the $\Gamma$ point, starting from the ground-state at T = 150 K, T = 300 K, and T = 450 K. 3 ps long runs were performed after an initial equilibration time of 1 ps with a 1 fs time-step. From heating at constant temperature a hexagonal two-dimensional slab composed of 125 atoms for 5 ps, no structural rearrangement of the Cr and B atoms was observed. Figure\ref{figMD} shows the average Cr-B, Cr-Cr, and intra-plane B-B bond length of monolayered CrB$_4$ at each temperature. No large variation of the inter-atomic distance between pairs of atoms demonstrate that the structural integrity of the nano-structure is preserved beyond the harmonic approximation, further confirming that the compound is stable at room temperature.

The electronic band structure of CrB$_4$ (Fig.\ref{fig2}b) features cone-shaped valence and conduction bands with a noteworthy difference with respect to other 2D semimetals: they meet at two different types of k-points in the irreducible BZ (iBZ), namely the $K$ (and equivalent $K'$) points, similarly to graphene, and on the line joining two high-symmetry points of the iBZ, as observed in 2D materials with rectangular unit cell \cite{PhysRevB.93.035401,PhysRevB.93.241405}.
Additionally, multiple other Dirac cones are present in the band diagram at different energies. In the following, the two Dirac points in the vicinity of the Fermi level will be the object of detailed analysis due to their relevance for electronic applications.

\begin{figure}[htp]
  \centering
  \includegraphics[width=0.5  \textwidth]{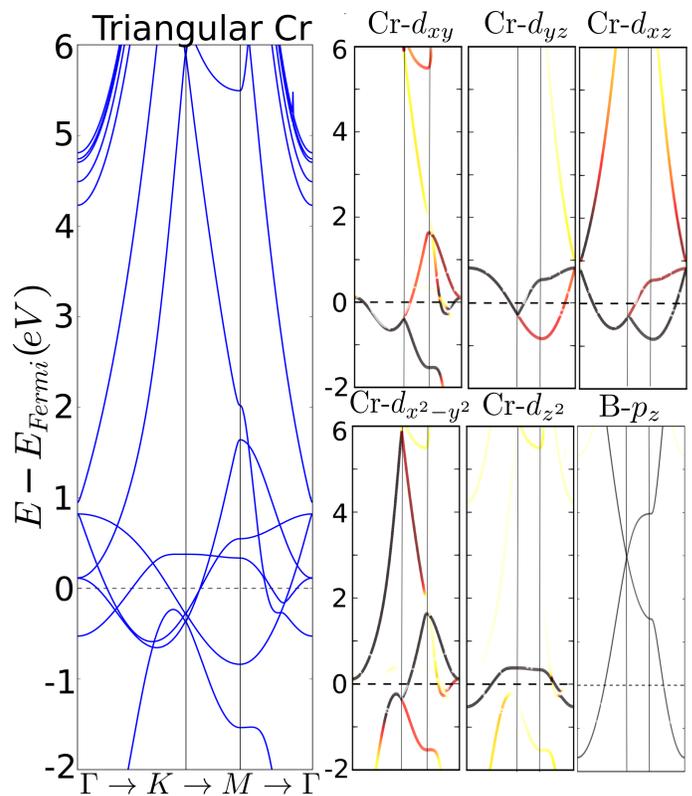}
 \caption{Left panel: Electronic band diagram of the triangular lattice composed by Cr atoms at the same positions of the CrB$_4$ lattice. The color-resolved panels correspond to the projected density of states of the d-orbitals, and show their independent contribution to each electronic state. Also the p$_z$ orbital of B atoms in the all-B honeycomb lattices is shown. }
 \label{fig3}
\end{figure}

To unveil the origin of each Dirac cone, we resort to a color scheme to represent the projected density of state (PDoS) diagrams, which allows for plotting the independent contribution of each atomic orbital to the electronic states (see Fig.\ref{fig2}). At $K$ point two band-crossings at different energy locations are worth noting. Slightly above the Fermi energy, the two perpendicular to the plane $d_{yz}$ and $d_{xz}$ orbitals combine to generate two bands that touch each other forming a Dirac point. The B $p_z$ orbitals also contribute to delocalize electrons over an energy range of $\sim$3 eV. The excess of holes created by this cone is compensated by another band-crossing forming a second Dirac point below the Fermi level along the $\Gamma-K$ line.

From the conduction to the valence band, the in-plane $d_{xy}$ derived band runs down $\sim$4 eV from the $\Gamma$ to the $K$ point, intersecting in a Dirac point with the coplanar $d_{x^2-y^2}$ and the perpendicular to the plane $d_{z^2}$ derived bands. 
Note that the contribution of an orbital determines the presence of a derived electronic band in the diagram, but the color intensity determines its degree of participation on the formation of that state. Thus, according to the color strength, the contribution of the B $p_z$ orbital to the second Dirac cone is comparatively less pronounced than in the former at the $K$ point. It can therefore be assumed that $p$-$d$ orbital hybridization is stronger at the $K$ point, as it is also suggested by the more pure Cr $d$-orbital character owned by the cone along the $\Gamma - K$ line.

\begin{figure}[htp]
 \centering
	 \includegraphics[width=0.5  \textwidth]{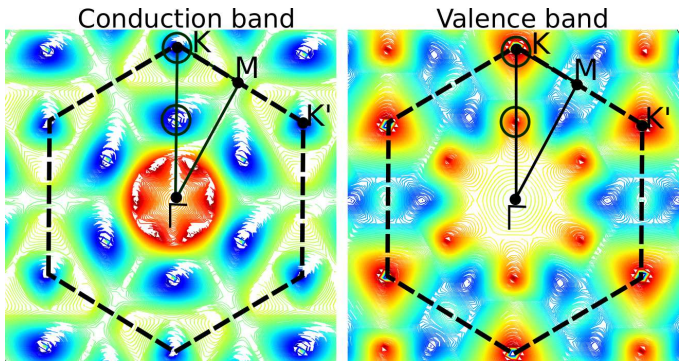}
 \caption{Constant-energy contour plots of the conduction and valence bands of CrB$_4$ in the first (delimited by dashed hexagon) Brillouin zone (BZ). $\Gamma$, M, and K (K') points delimit the irreducible BZ. By rotating 60$^\circ$ around $\Gamma$, a total of 12 Dirac cones in each band are obtained, six inequivalent along the $\Gamma - K$ and $\Gamma - K'$ lines, and six threefold degenerate (shared with neighbouring BZ) at $K$ and $K'$ points. Circles point out the location of the CrB$_4$ Dirac points.}
 \label{fig4}
\end{figure}

To additionally support this picture, the band diagram of the triangular lattice formed solely by Cr atoms at fixed positions is plotted in Fig.\ref{fig3}. The color-resolved PDoS of this lattice is displayed along with the one of B atom $p_z$ orbitals in the all-B monolayers. With respect to the CrB$_4$ monolayer, the dispersion in energy of the $d_{xy}$ band is reduced to a smaller range in the $\Gamma - K$ line. $d_{yz}$ and $d_{xz}$ create a Dirac point below the Fermi level and exhibit a strong hybridization along the $K - M - \Gamma$ path.  
A second band crossing slightly below the former point is formed by the $d_{xy}$ and the $d_{x^2-y^2}$ derived bands. Unlike in the hybrid structure, the latter is absent along the G-K line at the Fermi level.
When B lattices are present, the first set of bands shifts up to form the hole-pocket, whereas the second group shifts down to $\sim$-1.1 eV to form the electron pocket. $d_{z^2}$ stays at the same location in the spectrum for both the hexagonal and the triangular lattices, although its dispersion is lower in the latter. Interestingly, the Dirac cone of the B $p_z$ orbital in the standalone honeycombed lattice is located at the same energy (3.0 eV) that in the hybrid structure, exhibiting a similar energy dispersion. 
Note that the B $p_z$ orbitals hybridize with practically all the Cr $d$-states, yielding an increase of the $d$-bands dispersion, and a shifting of their location in the energy spectrum. 

\begin{figure}[htp]
 \centering
	 \includegraphics[width=0.5  \textwidth]{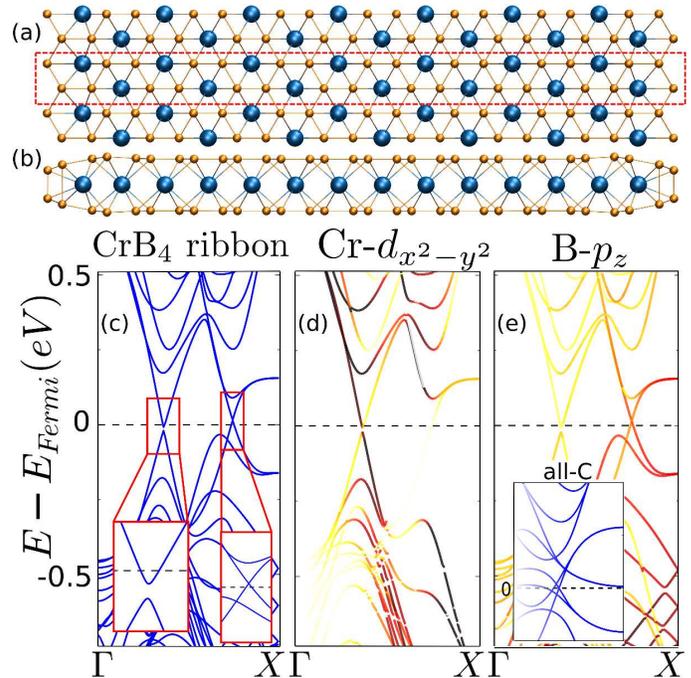}
 \caption{a) Top view of a CrB$_4$ ribbon with B-terminated zigzagged edges. Dotted rectangle delimits the unit cell of the ribbon. b) View along the ribbon axis of the same structure in a). c) Electronic band diagram of the CrB$_4$ ribbon. One frustrated Dirac cone is observed in the first half of the one-dimensional irreducible Brillouin zone, while two different bands form a Dirac cone in the second half. Zoom-in insets show the compensated electron-hole pocket. d) displays the contribution of the $d_{x^2-y^2}$ orbitals of all Cr atoms in the ribbon to the band diagram of a). Similarly, e) shows the contribution of all B atoms $p_z$ orbitals. Inset shows the band diagram close to the zone edge of an equivalent all-C tubular ribbon where the Dirac cone at Fermi level is also present.}
 \label{fig5}
\end{figure}

It is worth remembering that graphene exhibits six degenerate Dirac cones at the $K$ and $K'$ points of the BZ, which are threefold degenerate since they are shared by three contiguous BZ. Applying the symmetry operations of the {\it Amm2} point group to the iBZ (rotation around $\Gamma$ point), the 2D CrB$_4$ sheet exhibits by pairs two different groups of six Dirac cones apiece. Fig.\ref{fig4} shows contour plots of both the conduction and the valence bands of the first BZ (pointed out by the dashed hexagons). 
In addition to the threefold degenerate Dirac cones at the $K$ points, a group of six inequivalent cones lying on the $\Gamma - K$ lines completes a total of twelve pairs of Dirac cones in the vicinity of the Fermi level. 

Due to the large size of the Cr atom, the non-relativistic band structure of monolayered CrB$_4$ is susceptible to undergoing some changes when SOC is invoked. 
Fig.\ref{fig2}c shows that CrB$_4$ remains metallic although meV-large band gaps at each set of Dirac cones are developed in the vicinity of the Fermi level. The cones lying on the $\Gamma - K$ line and on the $K$ point exhibit a 14 meV-large gap and a 21 meV-large band gap respectively. Therefore, the major effect of the SOC is to vanish the Dirac points by slightly shifting the Dirac cones in energy. 

Patterning a 2D sheet into ribbons of small lateral size is an extended experimental practice to add new functionalities to a material. Lithographic methods \cite{Berger1191} and self-organized growth \cite{ISI:000282578000011} have demonstrated great success in the synthesis of graphene ribbons with defined edge geometries \cite{PhysRevLett.98.206805}. Self-assembly of boron atoms on Ag(110) surface has been reported as a controllable fabrication method of uniform all-B ribbons \cite{PhysRevMaterials.1.021001}. In the following, ribbons of CrB$_4$ of several widths are analyzed, showing that a Dirac cone is featured for a well-defined edge orientation.

Fig.\ref{fig5}a shows a CrB$_4$ ribbon unit cell as patterned along the direction shown in Fig.\ref{fig1}a. This particular ribbon unit cell is 36.5 \AA\ wide and 2.89 \AA\ long. B (Cr) atoms are arranged in an armchair (zigzagged) orientation across the ribbon width. Edges are terminated with B atoms in a zigzagged geometry. Distance between two Cr atoms in the unit cell progressively decreases from 2.92 \AA\ for inner atoms down to 2.68 \AA\ for the two outermost Cr atoms at the edges. Bond length between B atoms elongates from 1.68 \AA\ up to 1.77 \AA\ at the edges, where parallel honeycombed lattices meet with a vertical B-B bond length of 1.64 \AA\ (see Fig.\ref{fig5}b). 
  
Fig.\ref{fig5}c displays the electronic band diagram of the ribbon unit cell shown in the models drawn above. In the first half of the 1D iBZ, an inset shows two cone-shaped bands separated by a 20 meV gap that lies below the Fermi level. The compensated hole pocket is observed in the second inset, where two bands touch each other slightly above the Fermi level forming a Dirac cone near the zone edge. Considering the whole 1D BZ, this type of CrB$_4$ ribbon owns two actual and two frustrated Dirac points. Including SOC in the band structure calculation barely modifies the non-relativistic results: the Dirac cone remains intact and a tiny 4 meV band-splitting occurs in the gapped cone-shaped bands. 

The most intriguing feature of this type of quasi-1D nanostructure is precisely the appearance of a Dirac point at Fermi level, which is observed for all ribbons analyzed with widths ranging from 21.3 \AA\ to 56.7 \AA. 
If hosted in a 1D structure, Dirac cones derive from specific requirements on the parent material's BZ folding \cite{RMProche} or its topological nature\cite{doi:10.1021/nl500206u}. Thus, rolling a graphene ribbon up in a carbon nanotube (CNT) allows boundary conditions along the tube circumference to be established. Wave vectors around the CNT are quantized and can take only a set of discrete values. Whenever the allowed k-vectors include the $K$ point, which is always attained in armchair CNTs, the low-energy spectrum of the CNT hosts Dirac cones. Also, isolated Dirac cone edge states were predicted on topological insulators of binary composition with the Dirac point located in the middle of the 2D bulk gap  \cite{doi:10.1021/nl500206u}. 
The absence of topological insulating features in CrB$_4$ points out to boundary conditions created upon the parallel boron layers establish a direct connection through B-B bonds at the edges as a plausible explanation for the formation of the Dirac cone. 

Colored PDoS panels displayed in Fig.\ref{fig5}c  provide an insight on the orbitals that contribute more significantly to the electronic states of the CrB$_4$ ribbon. The different origin of the two sets of cone-shaped bands is evident: while the Dirac point is formed by the overlap of B $p_z$ orbitals, in-plane Cr $d_{x^2-y^2}$ orbitals generate the two bands which, for all ribbon widths, never close the gap. This result pinpoints to enhanced symmetry conditions in the case of $p$ orbitals as to forming the Dirac point, and a lack of symmetry on the $d$-orbitals as responsible for the gap. Indeed, while the triangular lattice of Cr atoms is infinite only the dimension of the ribbon axis, the two honeycombed lattices connected at the ribbon edges form a short of flat boron nanotube. The four electrons yielded by a Cr atom to each of the four B atoms allow the B sublattice to be isoelectronic to an only-C tubular ribbon. 

This interpretation is further supported by the electronic band diagram shown in inset of Fig.\ref{fig5}e of an isomorph structure to the CrB$_4$ ribbon where B were substituted by C atoms, and Cr atoms were removed. The nano-structure is equivalent to a flat-geometry armchair CNT. The Dirac cone close to the ribbon zone edge is similar to the cone found in CrB$_4$. Two groups of two degenerate bands at $\sim\pm$0.20 eV at the $X$ point develop down and up in energy and become non-degenerate as they approach the Fermi level, where two of them meet and form a Dirac cone. 

The exact length of the ribbon unit cell depends on its width, and it is smaller for narrower ribbons, while it tends to the bulk parameter as the ribbon width increases. 
Concurrently, the gap between the cone-shaped bands is highly sensitive to the size of the unit cell, namely to applied external strain. For compressive strain (or narrow ribbons) the electron pocket may switch to hole pocket, and vice-versa for tensile strain (or wide ribbons). For narrow ribbons (and some critical cell parameter of the wider ones), the ribbon Fermi surface is reduced to the Dirac point. No Peierls instability and band gap opening is observed upon doubling the ribbon unit cell.

\section{Conclusions}
  
In 2D, the canonical cones of graphene, which are derived from two symmetrically equivalent sublattices composed of just a single C atom apiece, have been reproduced by other materials of the same IV group, and rarely by compounds containing more than one element.
Here, it has been demonstrated that the C-free binary compound CrB$_4$ containing $p$- and $d$-orbitals can generate massless Dirac states with linear dispersion in both one- and two-dimensional geometries. As a result of the interplay between the Cr and B sublattices, CrB$_4$ hosts twelve pairs of Dirac cones in the vicinity of Fermi level (twice as many graphene) distributed equally at the six $K$ and $K'$ points and along the six $\Gamma - K$ lines of the BZ. When ribbons are formed by reducing CrB$_4$ dimensionality from 2D to quasi-1D, the parallel B lattices meet in zigzagged edges and boundary conditions are created. The resulting structure is topologically equivalent to flattened armchair CNTs, exhibiting a Dirac point which is independent of the ribbon contour-length. Furthermore, this Dirac state is practically decoupled of the Cr atoms sublattice, which due to the lack of symmetry only exhibits a frustrated Dirac point. It can be concluded that the extraordinary amount of Dirac cones in monolayered CrB$_4$ is the result of the presence of two types of atoms that actively participate and mutually contribute to the formation of the cones. This is possible due to the charge transfer from Cr $d$- to the B $p_z$-orbitals, which stabilizes both sublattices, shifts the crossing bands to the Fermi level, and increases their dispersion. Once the foundations for the creation of multiple Dirac states have been presented, numerous other materials with similar properties to monolayered CrB$_4$ can be predicted. Following the same theoretical strategy presented in this paper, free-standing WB$_4$ is suggested as a potential candidate to exhibit dynamical stability and Dirac states dominated by relativistic effects due to an enhanced SOC. Finally, experimental results on all-B 2D materials showed that Dirac cones are preserved despite the overlayer-substrate interactions \cite{PhysRevLett.118.096401}, suggesting that the quantum states predicted in this study could be preserved even when in interaction with a substrate. 

\section{Acknowledgments} 
This work was supported by the U.S. Department of Energy. I acknowledge the computing resources provided on Blues and Bebop, the high-performance computing clusters operated by the Laboratory Computing Resource Center at Argonne National Laboratory. 

\bibliography{biblio}{1}  
\end{document}